\documentclass[11pt,aps,showpacs,showkeys,preprintnumbers,amsmath,amssymb,nofootinbib]{revtex4}

\usepackage{subfigure}
\usepackage{graphicx}
\usepackage{color}
\usepackage{amsmath}
\usepackage{amsfonts}
\usepackage{amssymb}
\usepackage{adjustbox}
\def \be  {\begin{equation}}
\def \ee  {\end{equation}}
\def \ee  {\end{equation}}
\def \bea {\begin{eqnarray}}
\def \eea {\end{eqnarray}}

\def \k   {\kappa}

\DeclareMathOperator{\sech}{sech}
\begin{document}

\preprint{ECTP-2019-09}
\preprint{WLCAPP-2019-09}

\title{Multiplicity per rapidity in Carruthers and hadron resonance gas approaches}

\author{Abdel Nasser Tawfik}
\email{atawfik@nu.edu.eg}
\affiliation{Nile University - Egyptian Center for Theoretical Physics (ECTP), Juhayna Square off 26th-July-Corridor, 12588 Giza, Egypt}

\author{Mahmoud Hanafy}
%\email{mahmoud.nasar@fsc.bu.edu.eg}
\affiliation{Physics Department, Faculty of Science, Benha University, 13518, Benha, Egypt}
\affiliation{World Laboratory for Cosmology And Particle Physics (WLCAPP), 11571 Cairo, Egypt}

\author{Werner Scheinast}
%\email{werner@jinr.ru}
\affiliation{Joint Institute for Nuclear Research - Veksler and Baldin Laboratory of High Energy Physics, Moscow Region, 141980 Dubna, Russia}

\date{\today}

\begin{abstract}  

The multiplicity per rapidity of the well-identified particles $\pi^{-}$, $\pi^{+}$, $k^{-}$, $k^{+}$, $\bar{p}$, $p$, and $p-\bar{p}$ measured in different high-energy experiments, at energies ranging from $6.3$ to $5500~$GeV, are successfully compared with the Cosmic Ray Monte Carlo (CRMC) event generator. For these rapidity distributions, we introduce a theoretical approach based on fluctuations and correlations (Carruthers) and another one based on statistical thermal assumptions (hadron resonance gas model). Both approaches are fitted to the two sets of results deduced from experiments and simulations. We found that the Carruthers approach reproduces well the full range of multiplicity per rapidity for all produced particles, at the various energies, while the HRG approach fairly describes the results within a narrower rapidity-range. While the Carruthers approach seems to match well with the Gaussian normal distribution, ingredients such as flow and interactions should be first incorporated in the HRG approach.

\end{abstract}. 


\maketitle

%%%%%%%%%%%%%%%%%%%%%%%%%%%%%%%%%%%%%%
%%%   Section I
%%%%%%%%%%%%%%%%%%%%%%%%%%%%%%%%%%%%%%
  
\section{Introduction}

Characterizing the particle production is seen as one of the main gaols of the relativistic heavy-ion experiments \cite{Tawfik:2014eba}. By investigating the properties of the different produced particles throughout the various stages of the nuclear collision, essential information on partnoic such as quark gloun-plasma (QGP) and on hadronic phases can be obtained \cite{Satz:2000bn,Muller:1994rb,Shuryak:1980tp,Cleymans:1986cq,Cleymans:1992zc,Kouno:1988bi}. The sophisticated nuclear process starting from deconfined QGP and going through phase transition into confined hadrons, or vice versa is - among athers - characterized by a final stage, at which the number of the produced particles is conjectured to be fixed (chemically frozen) \cite{Andronic:2012ut}. The information, which can be deduced from the different experiments such as the ones at the Super Protonsyncrotron (SPS) at CERN, the Relativistic Heavy Ion Collider (RHIC) at BNL, and the Large Hadron Collider (LHC) at CERN, on the characterization of the particle distributions and the dynamical evolution of such strongly interacting systems, for instance, elliptic and radial flow, manifests an essential property of the particle spectra \cite{Tiwari:2012jq}. Various quantities such as particle ratios, transverse mass spectra, and multiplicity per rapidity are also crucial in studying dynamics and different general properties, especially at the freezeout stage \cite{Srivastava:2012bv}.    

The rapidity distribution, for example, was studied in different approaches \cite{Hagedorn:1980kb,Marques:2015mwa,Tiwari:2011km,Bjorken:1982qr,Schnedermann:1992qy,Schnedermann:1993ws,BraunMunzinger:1995bp,BraunMunzinger:1994xr,Feng:2011zze,Feng:2009zzb,Uddin:2011bi,Hirano:2001yi,Morita:2002av,Manninen:2010yf,Mayer:1997mi,Broniowski:2001we,Broniowski:2001uk,Bozek:2007qt}. Assuming superposition of two fireballs along the rapidity axes, the Tsalis-type distribution was assumed to give a successful description for the rapidity distribution \cite{Marques:2015mwa}. A thermodynamically consistent excluded volume model, in which flow is included, was proposed to well reproduce the rapidity distribution \cite{Tiwari:2012jq,Tiwari:2011km}. A relative better description was obtained when longitudinal collective flow and excluded volume corrections have been taken into consideration \cite{Schnedermann:1992qy, Schnedermann:1993ws}. Impacts of the hydrodynamic flow have been introduced in refs. \cite{BraunMunzinger:1995bp,BraunMunzinger:1994xr}. In \cite{Feng:2011zze, Feng:2009zzb}. In a single statistical thermal freezeout model based on a single set of parameters, the rapidity distribution was fairly analyzed \cite{Uddin:2011bi}. Also, in a hydrodynamical model with single particle spectra, the rapidity distribution was studied \cite{Hirano:2001yi,Mayer:1997mi}. The rapidity distribution was estimated in hydrodynamic models \cite{Morita:2002av}. Extended statistical hadronization models have been utilized in describing the repidity distribution \cite{Manninen:2010yf}.  In \cite{Broniowski:2001we,Broniowski:2001uk}, an extended thermal model is used to calculate particle spectra, at RHIC energies. In ref. \cite{Bozek:2007qt}, the rapidity distribution was calculated assuming longitudinal hydrodynamic expansion of fluid created in the heavy-ion collisions.

In statistical models \cite{Becattini:2007qr,Cleymans:2007jj,Broniowski:2001we,Broniowski:2001uk,Bozek:2007qt,Tawfik:2019gpc}, the chemical potential $\mu$ was successfully related to the rapidity $y$. Accordingly, the rapidity distribuation could be calculated over a wide large range of rapidity \cite{Tawfik:2019gpc}. An extended longitudinal scalling was also introduced to the thermal model \cite{Cleymans:2007jj}. 

It is well known that the distributions of the various particles are not isotropic. Thus, it is obvious to conclude that the multiplicity per rapidity calculated within the hadron resonance gas (HRG) model should not convinsingly reproduce the experimental results, especially at a wide range of rapidity \cite{Tiwari:2011km}. The inclusion of heavy resonances with their decay channels likely improves the HRG reproduction of the measured rapidity multiplicities. The best HRG results are the ones within small-rapidity region. The present script presents other alternatives aiming at improving HRG \cite{Lee:2004bx,Afanasiev:2002mx,Bearden:2002ib,Biedron:2006vf,Bratkovskaya:2017gxq,Adams:2003ve,Zhou:2009zzh,Hebeler:2010xb,Bearden:2004yx,Seyboth:2005rn,Adler:2001aq,Adcox:2003nr}. We also confront our results to the predictions from the Cosmic Ray Monte Carlo (CRMC) event generator \cite{Kalmykov:1997te, Kalmykov:1993qe, Ostapchenko:2005nj, Ostapchenko:2004ss, Ostapchenko:2007qb,Engel:1992vf, Fletcher:1994bd, Ahn:2009wx, Werner:2005jf,Pierog:2009zt}, at energies ranging from $6.3$ up to $5500~$GeV. The CRMC event generator includes various hadronic interaction models such as EPOS1.99 and EPOSlhc models. In this context, We use EPOSlhc hadronic interaction model from high down to low energies. 

For rapidity distributions measured in high-energy collisions and/or simulated in CRMC, we propose the utilization of Carruthers approach, which is based on hierarchy of cumulant correlation functions and their linked-pair approximation. This approach assumes an approximate translation invariance and utilizes a linking of averaged factorial moments to the second-order experimental moment in the final state of the particle production. For rapidity histograms, the various bins are assumed being irregular, i.e. they are influenced by fractal attractor or have an intermittent nature. The full range of rapidity $(\Delta y)^p$ can be then divided into smaller hypercubes of size $(\delta y)^p$ and thus the ordinary bin-averaged factorial moments can be determined, which - in turn - can be expressed in a linked-pair approximation. Relating this to the negative binomial distribution makes it possible to propose a specific functional form such as Gaussian or a general exponential for the cumulants.

The present paper is organized as follows. We briefly introduce Carruthers and HRG approaches in section \ref{models}. The results are discussed in section \ref{results}. The conclusions are outlined in section \ref{conc}.

\section{Models}
\label{models}

In this section, we give a brief description for the particle multiplicity as deduced from the HRG approach which is based on statistical thermal assumptions and the Carruthers approach which is based on correlations and fluctuations. 

\subsection{HRG approach for rapidity multiplicity} 
\label{sec:hrg}  

It is widely known that the formation of resonances could be understood as bootstrap, i.e. the fireballs or resonances are demonstrated to be consisting of more smaller fireballs or lighter resonances, which - in turn - are composed of smaller fireballs and lighter resonances etc. For such a system, the thermodynamics quantities can be derived directly from the partition function $Z(T, \mu, V)$.  In a Grand canonical ensemble, the partion function reads \cite{Tawfik:2014eba}
\begin{equation} 
Z(T,V,\mu)=\mbox{Tr}\; \left[\exp\left(\frac{{\mu}N-H}{T}\right)\right], \label{eq(1)}
\end{equation}
where $H$ is Hamiltonian of the system, $N$ is the total number of constituents. In the HRG model, Eq. (\ref{eq(1)}) can be expressed as a sum over all hadron resonances\footnote{either compiled by the particle data group \cite{Tanabashi:2018oca} or still theoretical predictions \cite{Capstick:1986bm}.} 
\begin{equation} 
\ln  Z(T,V,\mu)=\sum_i{{\ln Z}_i(T,V,\mu)} =\frac{V g_i}{(2{\pi})^3}\int^{\infty}_0{\pm  d^{3}p\; {\ln} {\left[1\pm \exp\left(\frac{{\varepsilon}_i(p)-\mu_{i}}{T} \right) \right]}}, \label{eq(2)}
\end{equation}
where $\pm $ stands for bosons and fermions, respectively and $\varepsilon_{i}=\left(p^{2}+m_{i}^{2}\right)^{1/2}$ is the dispersion relation of the $i$-th particle, for which the total number of particles can be obtained as
\begin{equation}
N_{i}=T \frac{\partial Z_{i}(T, V)}{\partial \mu_{i}}=\frac{V g_i}{(2 \pi)^3}\; \int^{\infty}_0 d^3 p\; \left[\exp\left(\frac{\varepsilon_{i} (p)-\mu_{i}}{T}\right) \pm 1 \right]^{-1}. \label{eq(3)} 
\end{equation}
From Eq. (\ref{eq(3)}), the invariant momentum spectrum of particles emitted from a thermal source can be derived \cite{Tawfik:2019oct,Tawfik:2019wze}
\begin{equation}
\varepsilon_{i} \frac{d^{3}N_{i}}{dym_{\mathtt{T}}dm_{\mathtt{T}}d\phi}=\varepsilon_{i} \frac{V g_i}{(2{\pi})^3}\left[\exp\left(\frac{\varepsilon_{i} (p)-\mu_{i}}{T}\right)\pm1\right]^{\mathtt{-1}},  \label{eq(4)}
\end{equation}
where $m_{\mathtt{T}}$ is the transverse mass and is given by $m_{\mathtt{T}}=\sqrt{m^{\mathtt{2}}+p_{\mathtt{T}}^{\mathtt{2}}}$, where $p_{\mathtt{T}}$ is the transverse momentum. The energy of the $i$-th particle, $\varepsilon_{i}$, can be then expressed in terms of the rapidity $\left(y\right)$ and $m_{\mathtt{T}}$ as $\varepsilon=m_{\mathtt{T}} \cosh\left(y\right)$. Through integration over the full transverse mass $m_{\mathtt{T}}$, the multiplicity $N$ per rapidity $y$ can be derived,
\begin{equation}
\frac{dN}{dy}=\sum_i \frac{V g_i}{\left(2{\pi}\right)^2}\int^{\infty}_0{ \cosh\left(y\right) m_{\mathtt{T}}^{2} \left[\exp\left(\frac{m_{\mathtt{T}} \cosh\left(y\right)-\mu_{i}}{T}\right)\pm1\right]^{-1}}.   \label{eq(7)}
\end{equation}
Eq. (\ref{eq(7)}) can be rewritten as
\begin{equation}
\frac{dN}{dy}=\sum_i \frac{V g_\mathtt{i}}{\left(2 {\pi}\right)^2} T \left[m_{\mathtt{T}}+2T\sech\left(y\right)\left(m_{\mathtt{T}}+T\sech\left(y\right)\right)\right]\exp\left[-\frac{m_{\mathtt{T}}\cosh\left(y\right)-\left(\mu_{i}\pm1\right)}{T}\right],    \label{eq(15)}
 \end{equation}
 where $T$ is the freezeout temperature, $\mu_{\mathtt{i}}$ is the chemical potential which can be related to the beam energy \cite{Tawfik:2013bza,Tawfik:2014eba}, $g_{\mathtt{i}}$ is the degeneracy, and $V$ is the volume of the fireball.  Eq. (\ref{eq(15)}) gives the multiplicity per rapidity for hadron states, whereas combination of the trigonometric functions $\sech$ and $\cosh$ are conjectured to assure  Gaussian-like distribution function. In classical statistics, Eq. (\ref{eq(15)}) can be reexpressed as  
 \begin{equation}
\frac{dN}{dy}=\sum_i \frac{T V g_i}{\left(4{\pi}\right)^2}\left(2T^{2}+m_{\mathtt{T}}\cosh\left(y\right)\left(2T+m_{\mathtt{T}}\cosh\left(y\right)\right)\right)\sech\left(y^{2}\right)\exp\left(\frac{\mu_{i}-m_{\mathtt{T}}\cosh\left(y\right)}{T}\right).     \label{eq.(cccc)} 
\end{equation} 

\subsection{Carruthers approach for rapidity multiplicity}
\label{sec:newmodel} 

Suppose we have a histogram for rapidity, $y$, describing an event, $l$, in the rapidity interval $\Delta y$, which is divided into $X$ bins of length $\delta y$. Consequently, the rapidity interval reads $\Delta y=X \delta y$. In the thermal models, the rapidity density of $i$-th particle is given as two-particle correlation integral  \cite{p.car}
\begin{equation}
\frac{dN^{l}\left(y\right)}{dy}= \int_{\delta y}{\rho^{\prime l}\left(y\right)}dy, \label{eq(aa)}
\end{equation}
where $\rho^{\prime l}\left(y\right)$ is the probability density corresponding to the considered regions ({\it fireballs}) of the latest (final) hadron yields. This probability density $\rho^{\prime l}\left(y\right)$ can be expressed as \cite{p.car}
\begin{equation} 
\rho^{\prime l}\left(y\right)=\sum_{\mathtt{i}} \delta \left( y- y^{l}_{\mathtt{i}}\right),     \label{eq(c)}
\end{equation}
in which $\delta \left( y- y^{l}_{i}\right)$ counts the avaliable number of points $y^{l}_{\mathtt{i}}$ in the known interval $\delta y$ for $l$-th event, i.e. $\delta$-function defines the rapidity bin width. For many events, the single particle density $dN/dy$ related to the differential cross-section $d\sigma/dy$ reads \cite{p.car}
\begin{equation}
\frac{1}{\sigma_{\mathtt{i}}}\frac{d\sigma}{dy}=\frac{1}{N^{l}}\sum_{\mathtt{l}}\frac{dN^{l}}{dy} \label{eq(dd)}
\end{equation}
where $N^{l}$ is the total number of particles and $\sigma_{i}$ is the corresponding cross-section.
This can be converted into a statistical ensemble having a probability density $\rho(y_{1},y_{2},\cdots,y_{N})$ for the distributed points $y_{\mathtt{i}}$, where $i=1,2,\cdots,N$ stand for particles within th interval $\Delta y$. Thus, Eq. (\ref{eq(dd)}) can be rewritten as \cite{p.car} 
\begin{eqnarray}
\label{eq(g)}
\frac{1}{\sigma_{i}}\frac{d\sigma}{dy}\equiv\rho_{1}\left(y\right)&=&\left\langle\sum_{\mathtt{i}} \delta \left(y-y_{\mathtt{i}}\right)\right\rangle, \nonumber   \\
\frac{1}{\sigma_{\mathtt{i}}}\frac{d^{2}\sigma}{dy_{1}dy_{2}}\equiv\rho_{2}\left(y_{\mathtt{1}},y_{\mathtt{2}}\right)&=&\left\langle\sum^{\prime}_{\mathtt{i},\mathtt{j}}\delta \left(y_{\mathtt{1}}-y_{\mathtt{i}}\right)\delta \left(y_{\mathtt{2}}- y_{\mathtt{j}}\right) \right\rangle, \\
\frac{1}{\sigma_{i}}\frac{d^{3}\sigma}{dy_{1}dy_{2}dy_{3}}\equiv\rho_{3}\left(y_{\mathtt{1}},y_{\mathtt{2}},y_{\mathtt{3}}\right)&=&\left\langle\sum^{\prime}_{i,j,k}\delta \left(y_{\mathtt{1}} - y_{\mathtt{i}}\right)\delta \left(y_{\mathtt{2}} - y_{\mathtt{j}}\right)  \left(y_{\mathtt{3}} - y_{\mathtt{k}}\right)\right\rangle,   \nonumber
\end{eqnarray} 
where $\rho_{q}$ are the rapidity distribution correlation functions and the hat to the summation refers to exclusion of terms with equal indices. Eqs. (\ref{eq(g)}) enable simplified calculations for the integrating the correlations from the fluctuations of the particle multiplicity considering different domains in the rapidity ranges. Assuming a domain $Q_{2}$ with equal ranges $\Delta y$ of $y_{i}$, thus Eq. (\ref{eq(aa)}) can be rewritten as follows \cite{1991IJMPA...6.3031C}.
\begin{eqnarray}
\label{eq(h)}
\int_{Q_{\mathtt{1}}}\rho_{\mathtt{1}}\left(y_{1}\right)dy_{\mathtt{1}}&=&\left\langle N \right\rangle_{Q_{\mathtt{1}}}, \nonumber  \\ 
\int_{Q_{\mathtt{2}}}\rho_{\mathtt{2}}\left(y_{\mathtt{1}},y_{\mathtt{2}}\right)dy_{\mathtt{1}}dy_{\mathtt{2}}&=&\left\langle n\left(n-1\right) \right\rangle_{Q_{\mathtt{2}}}, \\
\int_{Q_{\mathtt{3}}}\rho_{\mathtt{3}}\left(y_{\mathtt{1}},y_{\mathtt{2}},y_{\mathtt{3}}\right)dy_{\mathtt{1}}dy_{\mathtt{2}}dy_{\mathtt{3}}&=&\left\langle n\left(n-1\right)\left(n-2\right) \right\rangle_{Q_{\mathtt{3}}}. \nonumber 
\end{eqnarray}
For application in high-energy analysis, the factorial moments of the averaged bin for $\rho=2$ can be rewritten as \cite{p.car}
\begin{equation}
\int_{Q_{\mathtt{2}}}\rho_{\mathtt{2}}\left(y_{\mathtt{1}},y_{\mathtt{2}}\right)dy_{\mathtt{1}}dy_{\mathtt{2}}=\sum^{X}_{k=1}\left\langle n_{k}\left(n_{k}-1\right) \right\rangle, \label{eq(k)}
\end{equation}
where $n_{k}$ is the number of particles in bin $\left(k\right)$. The summation over hypercubes in higher dimensions $Q_{q}=\sum\left(\delta y\right)^{n}$ gives (a generalization to $Q_{q}$):
\begin{equation}
\int_{Q_{q}}\rho_{q}\left(y_{\mathtt{1}},y_{2},\cdots,y_{q}\right)dy_{1}dy_{2} \cdots dy_{q}=\sum^{X}_{k=1}\left\langle n_{k}\left(n_{k}-1 \cdots \left(n_{k}-n-1\right)\right) \right\rangle. \label{eq(m)}
\end{equation}
When removing the symmetry of low-order density correlations, the cumulants can be used. 
The correlation functions of the cumulants $\left(c_{p}\right)$ are then to be expressed in terms of correlation densities and vice versa as the follows (number of permutations in the sums are shown in brackets) \cite{1991IJMPA...6.3031C}.
\begin{eqnarray}
 \rho_{2}\left(1,2\right)&=&\rho_{1}\left(1\right)\rho_{1}\left(2\right)+c_{2}\left(1,2\right), \nonumber\\ 
\rho_{3}\left(1,2,3\right)&=&\rho_{1}\left(1\right)\rho_{1}\left(2\right)\rho_{1}\left(3\right)+\sum_{\left(3\right)}c_{2}\left(1,2\right)\rho_{1}\left(3\right)+c_{3}\left(1,2,3\right), \label{eq(p)}\\
\rho_{4}\left(1,2,3,4\right)&=&\rho_{1}\left(1\right)\rho_{1}\left(2\right)\rho_{1}\left(3\right)\rho_{1}\left(4\right)+\sum_{\left(4\right)}\rho_{1}\left(1\right)\rho_{1}\left(2\right)c_{2}\left(3,4\right) \nonumber\\
&+&\sum_{\left(3\right)}c_{2}\left(1,2\right)c_{2}\left(3,4\right)+\sum_{\left(4\right)}c_{3}\left(1,2,3\right)\rho_{1}\left(4\right)+c_{4}\left(1,2,3,4\right). \nonumber
\end{eqnarray}
The factorial of cumulant moments $\left(f_{q}\right)$ reads \cite{p.car}
\begin{eqnarray}
f_{2}&=&\left\langle n\left(n-1\right) \right\rangle -\left\langle n \right\rangle ^{2}, \label{eq(s)}\\   
f_{3}&=&\left\langle n\left(n-1\right)\left(n-2\right)\right\rangle -3\left\langle  n\left(n-1\right) \right\rangle \left\langle n \right\rangle +2 \left\langle n \right\rangle ^{3},  
\end{eqnarray}
and so on, are just integrals of the corresponding cumulants $\left(c_{p}\right)$.

The moments can be then averaged over all bins, $X$ numbers of identical widthes $\left(\delta y \right)$ normalized either to the overall mean number per bin, $\left[ \bar{n}=\bar{\rho} \delta y =\sum^{X}_{m=1}\left\langle n_{m} \right\rangle /X \right]$ or to the local average $\left\langle n_{m}\right\rangle\equiv \bar{\rho_{m}}\delta y$. These choices are usually referred to as "horizontal" and "vertical" averages, respectively, \cite{p.car}
\begin{equation}
F^{h}_{q}\equiv\frac{1}{X\left(\delta y\right)^q}\sum^{X}_{m=1}\int_{Q_{m}}\prod_{i}dy_{i} \frac{\rho_{q}\left(y_{1} \cdots y_{q}\right)}{\bar{\rho^{q}}}, \label{eq(v)}
\end{equation}
\begin{equation}
F^{v}_{q}\equiv\frac{1}{X}\sum^{X}_{m=1}\frac{\left\langle n_{m}\left(n_{m}-1\right) \cdots \left(n_{m}-q+1\right)\right\rangle}{\left\langle n_{m} \right\rangle^{q}}. \label{eq(w)} 
\end{equation}
Recalling the factorial cumulants of the averaged bin
\begin{eqnarray}
K^{h}_{q}\left(\delta y\right) \equiv\frac{1}{X}\sum^{X}_{m=1}\frac{f^{m}_{q}}{\bar{n^{q}}}&=&\frac{1}{X\left(\delta y\right)^q}\sum^{X}_{m=1}\int_{Q_{m}}\prod_{i}dy_{i} \frac{c_{q}\left(y_{1} \cdots y_{q}\right)}{\bar{\rho^{q}}} =\sum_{m}\frac{N_{m}\left(N_{m}-1\right)}{2\sqrt{\pi}\sigma_{\mathtt{T}}}e^{-y^{2}/4\sigma^{2}_{\mathtt{T}}}, \label{eq(y)} \\ 
K^{v}_{q}\left(\delta y\right) \equiv\frac{1}{X}\sum^{X}_{m=1}\frac{f^{m}_{q}}{\left\langle n_{m} \right\rangle}&=&\frac{1}{X\left(\delta y\right)^q}\sum^{X}_{m=1}\int_{Q_{m}}\prod_{i}dy_{i} \frac{c_{q}\left(y_{1} \cdots y_{q}\right)}{\left\langle \bar{\rho_{m}}\right\rangle ^{q}}=\sum_{m\neq m\prime}\frac{N_{m}N_{m\prime}}{2\sqrt{\pi}\sigma_{\mathtt{T}}}e^{-\left(y-\bar{y}_{m}+\bar{y}_{m\prime}\right)^{2}/4\sigma^{2}_{\mathtt{T}}}. \label{eq(y2)} \hspace*{8mm}
\end{eqnarray}
It is obvious that when the values of one variable, such as $\left(y_{\mathtt{1}}\right)$ approaches zero, the dependence on $y_{\mathtt{2}}$ of the correlated particle could be fitted as an either exponential or Gaussian distribution in order to deduce the cumulant $K_{2}$ \cite{p.car}
\begin{equation}
K_{\mathtt{2}} \equiv \frac{\rho_{\mathtt{2}}\left(y_{\mathtt{1}},y_{\mathtt{2}}\right)-\rho_{\mathtt{1}}\left(y_{\mathtt{1}}\right)\rho_{\mathtt{1}}\left(y_{\mathtt{2}}\right)}{\rho_{\mathtt{1}}\left(y_{\mathtt{1}}\right)\rho_{\mathtt{1}}\left(y_{\mathtt{2}}\right)} \approx \gamma_{\mathtt{2}}e^{-\left(y_{\mathtt{1}}-y_{\mathtt{2}}\right)^{2}/4\lambda^{2}},  \label{eq(bb)}
\end{equation}
where the rapidity density of a source $n$ emitting $\int dy \rho_1^{(m)}(y)$ particles \cite{Randrup:2004fs}
\begin{equation}
\rho^{\left( n\right)}_{1}\approx\sum_{\mathtt{n}}\frac{N_{\mathtt{n}}}{\sqrt{2\pi}\sigma_{\mathtt{T}}}e^{-\frac{1}{2}\left(y-\bar{y}_{\mathtt{n}}\right)^{2}/\sigma^{2}_{\mathtt{T}}}, \label{eq(cc)}
\end{equation}
which can be directly related to $dN/dy$.

\section{Cosmic Ray Monte Carlo (CRMC) model}
\label{sec:crmc} 

As introduced, the hybrid Cosmic Ray Monte Carlo (CRMC EPOSlhc) event generator shall be utilized in generating various multiplicity per rapidity for different hadrons, at energies ranging between $\sqrt{s_{NN}}=6.3$ and $5500~$GeV. The CRMC EPOSlhc results are then compared with the available experimental data and finally fitted by the HRG and Carruthers approaches.

The CRMC is an interface for the various cosmic ray monte-carlo models for various effective quantum chrmodynamic (QCD) models and different experiments such as CMS, ATLAS, LHCb, NA61 and the ultra high-energy cosmic rays obervatory Pierre Auger, etc. It includes different types of interactions that are built on the highly Gribov-Regge model-like EPOSlhc/$1.99$. CRMC introduces a full description for background taking into consideration the resultant differaction. Its interface can access the resultant output from various event-generators for heavy-ion collisions. CRMC interface is also connected to a wide spectrum of models, such as, qgsjet$01$ \cite{Kalmykov:1997te, Kalmykov:1993qe}, qgsjetII \cite{Ostapchenko:2005nj, Ostapchenko:2004ss, Ostapchenko:2007qb}, sibyll \cite{Engel:1992vf, Fletcher:1994bd, Ahn:2009wx} and EPOS $1.99$/lhc \cite{Werner:2005jf, Pierog:2009zt}. QGSJET$01$ and SIBYLL$2.3$, at low energies. EPOS lhc/$1.99$, QGSJETII v$03$ and v$04$ are the interaction models that can be integrated at high energies.   

In the present paper, we utilize the CRMC EPOSlhc event generator which contains various parameters for the primordial observables in high-energy collisions and their phenomenological assumptions. These can be modified due to theoretical and experimental postulates. It was argued that EPOSlhc is able to give a reasonable description for heavy-ion collisions regarding the generated data from various experiments and aso other event generators \cite{Werner:2005jf, Pierog:2009zt}. 

EPOSlhc was originally constructed for cosmic ray air showers and could be utilized for pp- and AA-collisions, at SPS, RHIC, and LHC energies. EPOSlhc even uses a more simplified treatments for heavy-ion collisions at the last stage of their evolutions and thus can be applied for minimum bias in the interactions between hadrons in the nuclear collisions \cite{Pierog:2013ria}. EPOSlhc is a parton model with various parton-parton interactions resulting in various parton ladders and provides a good estimation for particle yields, multiple scattering of partons, evaluations of cross-section, shadowing and screening through splitting and unitarization, and various collective effects of hot and dense media. It should be mentioned that EPOSlhc does not consider the simulations for complete hydro system even in the last stage. 

In the present work, we utilize EPOSlhc event generator, at energies spanning between $6.3$ and $5500~$GeV for an ensemble of at least $100,00$ events. We have calculated the multiplicity  for $\pi^{-}$, $\pi^{+}$, $k^{-}$, $k^{+}$, $\bar{p}$, $p$, and $p-\bar{p}$ in various rapidity windows $-6 < y < 6$. Proving the validity of the hybrid EPOSlhc event generator, we hope at calculating the multiplicity per rapidity for the considered various hadrons, which is assumed to come up with a novel input for the future facilities NICA and FAIR, for instance.

\section{Results and Discussion}
\label{results}  
	
The present analysis is based on a reproduction of experiments results for the multiplicity per rapidity $dN/dy$ of  $\pi^{-}$, $\pi^{+}$, $k^{-}$, $k^{+}$, $\bar{p}$, $p$, and $p-\bar{p}$ \cite{Lee:2004bx,Afanasiev:2002mx,Bearden:2002ib,Bratkovskaya:2017gxq,Adams:2003ve,Zhou:2009zzh,Hebeler:2010xb,Bearden:2004yx,Seyboth:2005rn}, at energies ranging between $6.3$ and $5500~$GeV. We also compare with results deduced from EPOSlhc event generator \cite{Werner:2005jf, Pierog:2009zt}. Both sets of results are then confronted to the HRG thermal and the Carruthers rapidity approaches, in which the dependence of the freezeout temperature $T$ and the baryon-chemical potential $\mu_{\mathtt{B}}$ on the centre-of-mass energy $\sqrt{s_{\mathtt{NN}}}$ is taken from ref. \cite{Tawfik:2014eba}. 

The present work aims at updating the study of the multiplicity per rapidity in the HRG model. One of the improvements we are presenting here is the inclusion of various missing states to the well-know hadron states recently reported by the particle data group \cite{Tanabashi:2018oca}. The missing states are hadron resonances which are theoretically predicted \cite{Capstick:1986bm}, but not yet confirmed, experimentally. Various physical characteristics including masses, and other quantum numbers, etc. are  theoretically well known. It was conjectured that these states greatly contribute to the fluctuations and the correlations simulated in the recent lattice QCD calculations \cite{Bazavov:2014xya}. Best reproduction of fluctuations and the correlations are among the main motivations to add these hadron states to the HRG model \cite{ManLo:2016pgd}. Regardless the corresponding limitations, we intend it check whether the new hadron states contribute to the multiplicity per rapidity of  $\pi^{-}$, $\pi^{+}$, $k^{-}$, $k^{+}$, $\bar{p}$, $p$, and $p-\bar{p}$, as the missing states likely come up with additional degrees of freedom and certainly considerable decay channels which might affect the final number of particles produced. 

Also the present work introduces a new approach based on Carruthers proposal for hierarchy of cumulant correlation functions and their linked-pair approximation, which satisfactorily characterizes the galaxy correlation and successfully describes the central rapidity domain \cite{p.car}. The basic idea is an approximate translation invariance and a linking of averaged factorial moments to the second-order experimental moment in the final state of the particle production. As for rapidity histograms, it was assumed that the various bins are likely irregular, i.e. they are influenced by fractal attractor, e.g.  intermittence, where the full range of rapidity $(\Delta y)^p$ is divided into smaller hypercubes of size $(\delta y)^p$. An ordinary bin-averaged factorial moments can be determined, which can then be expressed in linked-pair approximation. This - in turn - can be related to the negative binomial distribution. For the high-energy collisions, a specific functional form such as Gaussian or an exponential can be proposed for the cumulants or the correlation functions, Eq. (\ref{eq(cc)}).

The results obtained are compared with both experiment and event generator. For almost all particles, the dependence of the multiplicity per rapidity $dN/dy$ on the rapidity $y$ was fitted to Gaussian normal distribution function,
\begin{equation}
\frac{d N}{d y}=a_{0} \exp\left\{-0.5\left[\left(\frac{y-a_{1}}{a_{2}}\right)\left(\frac{y-a_{1}}{a_{2}}\right)\right]\right\}, \label{eq.(llll)}
\end{equation}
where $a_{0}$, $a_{1}$, and $a_{2}$ are the fit parameters, Tab. \ref{tab1}.	For net proton $p-\bar{p}$, we use the binomial
\begin{equation}
\frac{d N}{d y}=c_{0}+c_{1}y+c_{2}y^{2}+c_{3}y^{3},     \label{eq.(ddd)}
\end{equation}
in which $c_{\mathtt{0}}$, $c_{\mathtt{1}}$, $c_{\mathtt{2}}$, and $c_{\mathtt{3}}$ are free parameters, Tab. \ref{tab2}.

\begin{figure}[!htb]
\includegraphics[width=7cm]{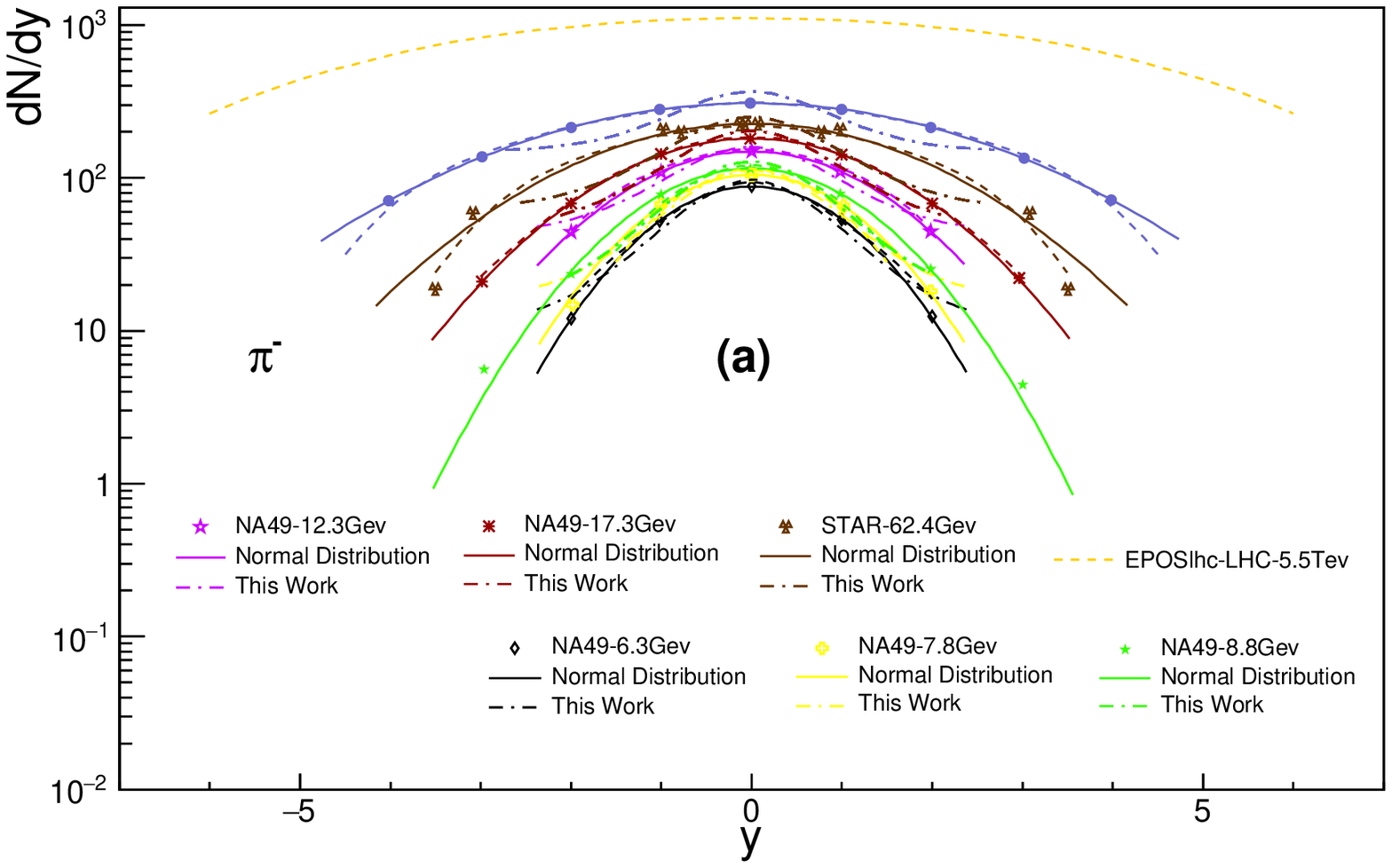}
\includegraphics[width=7cm]{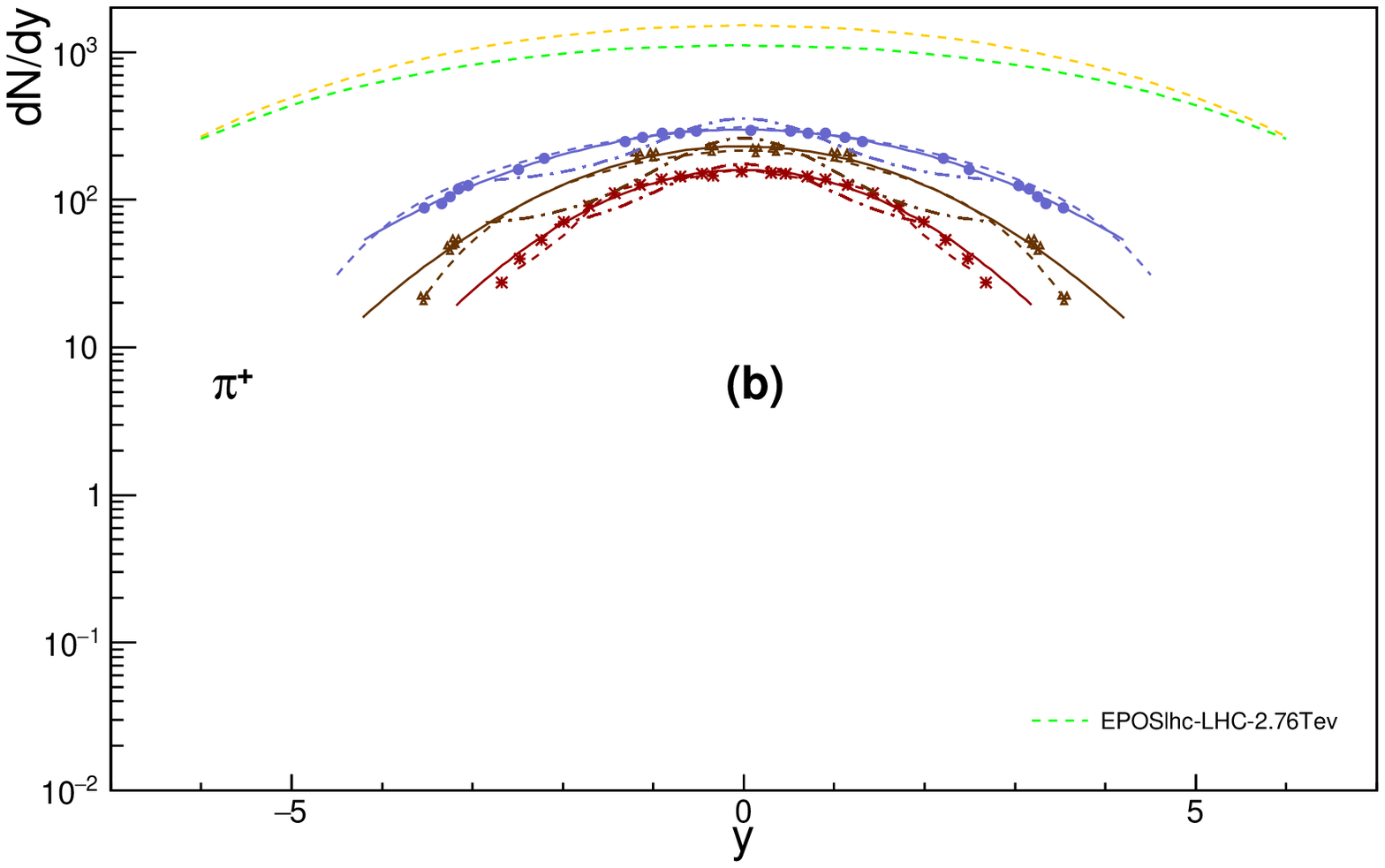} 
\includegraphics[width=7cm]{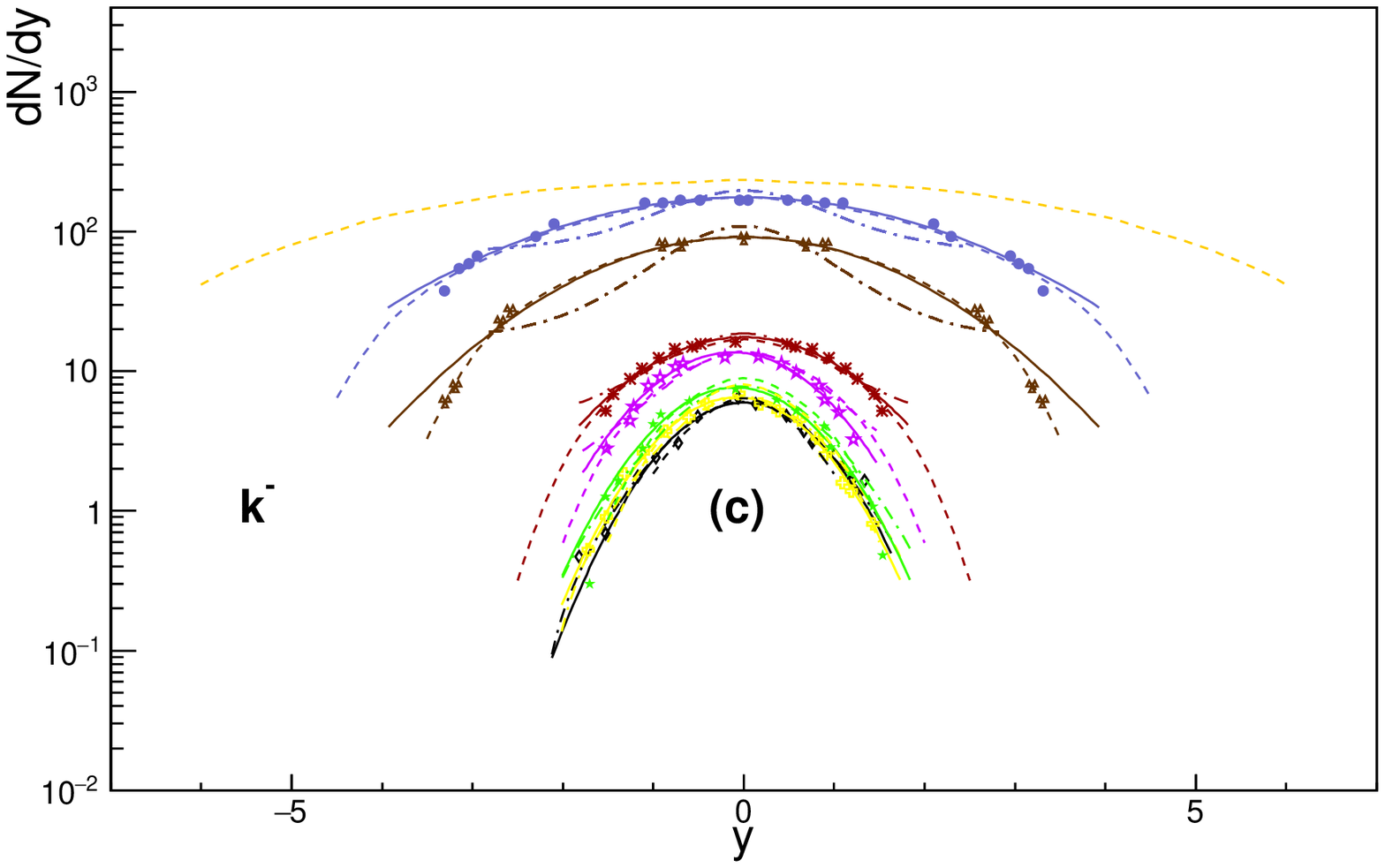} 
\includegraphics[width=7cm]{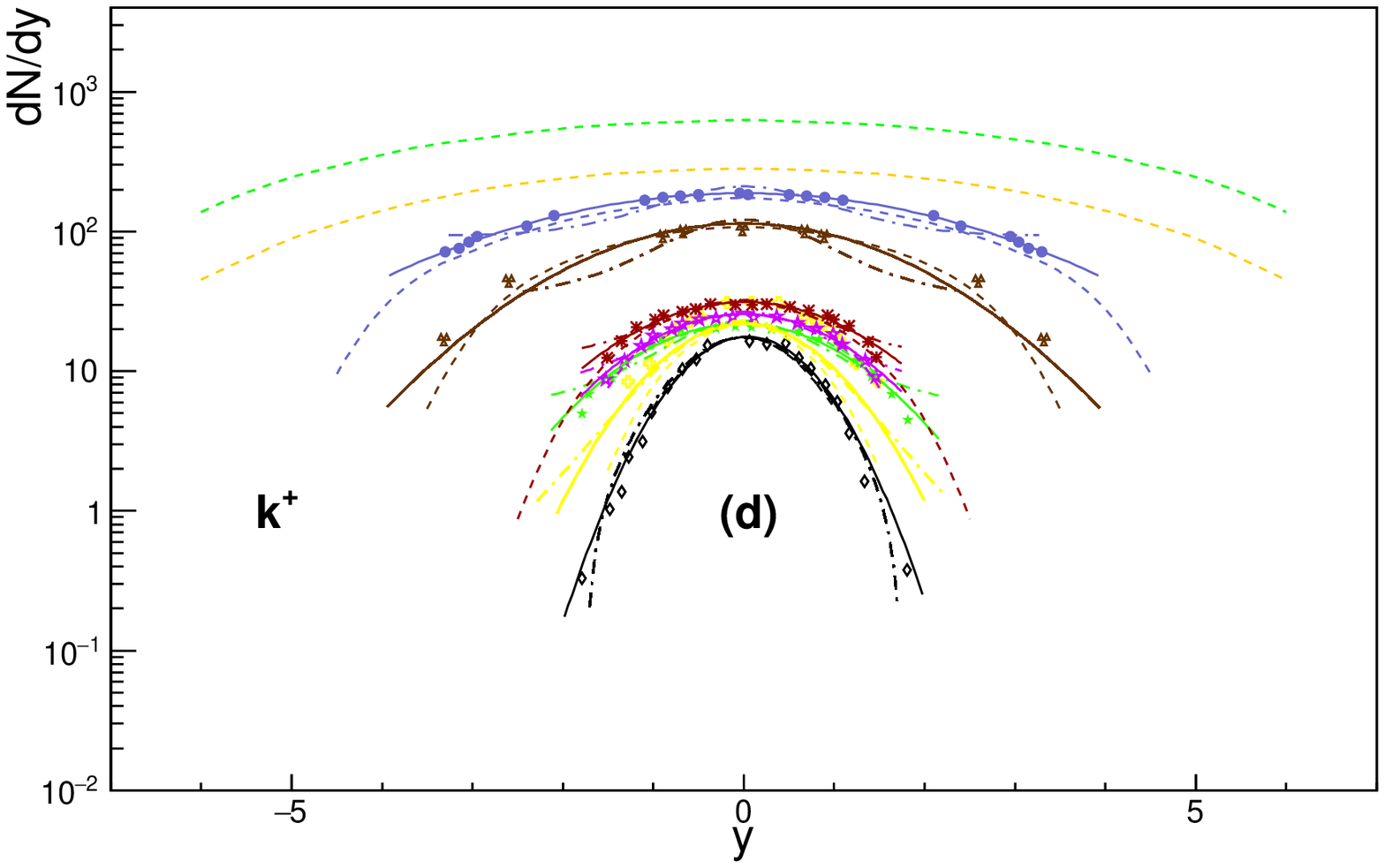}
\includegraphics[width=7cm]{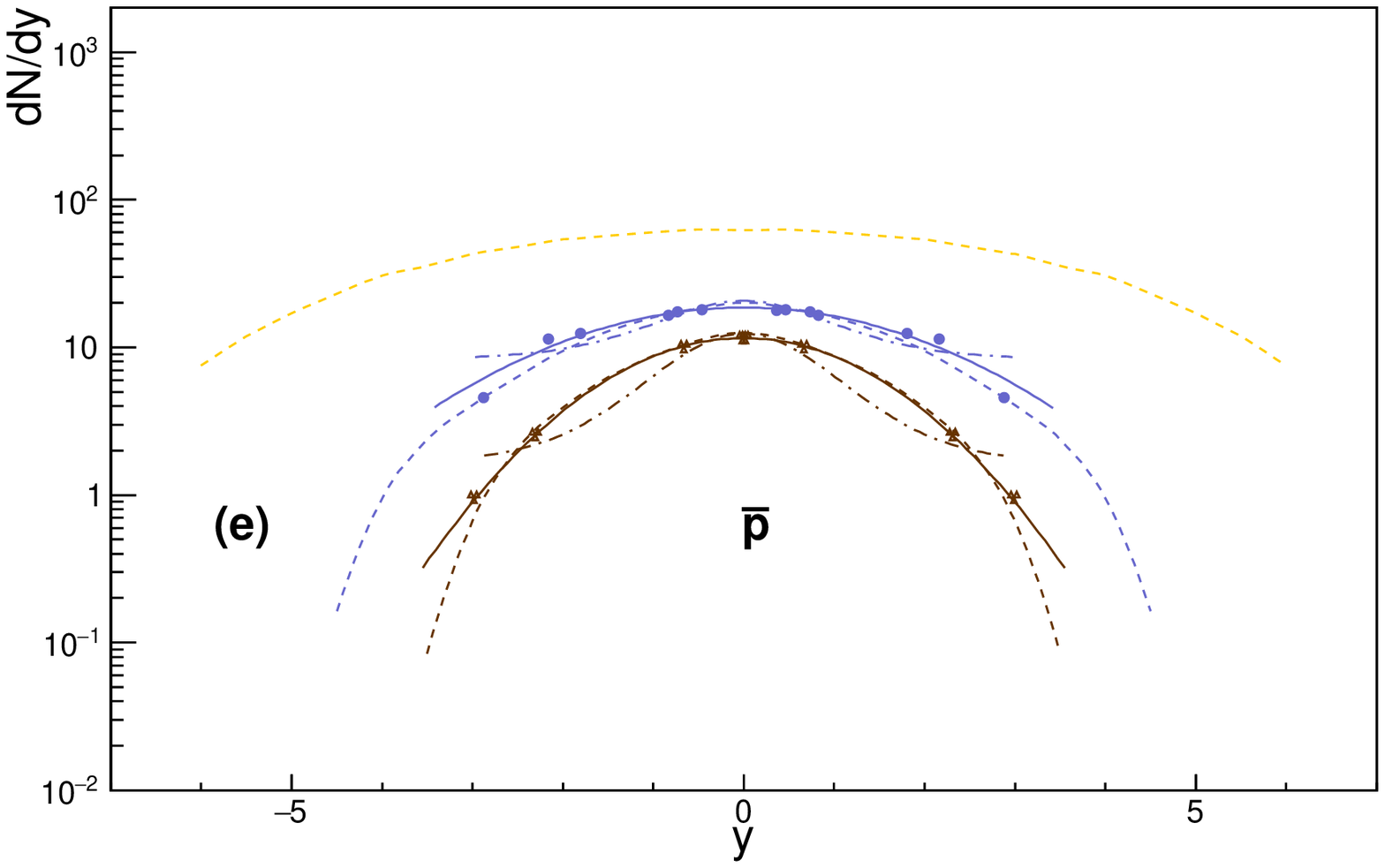} 
\includegraphics[width=7cm]{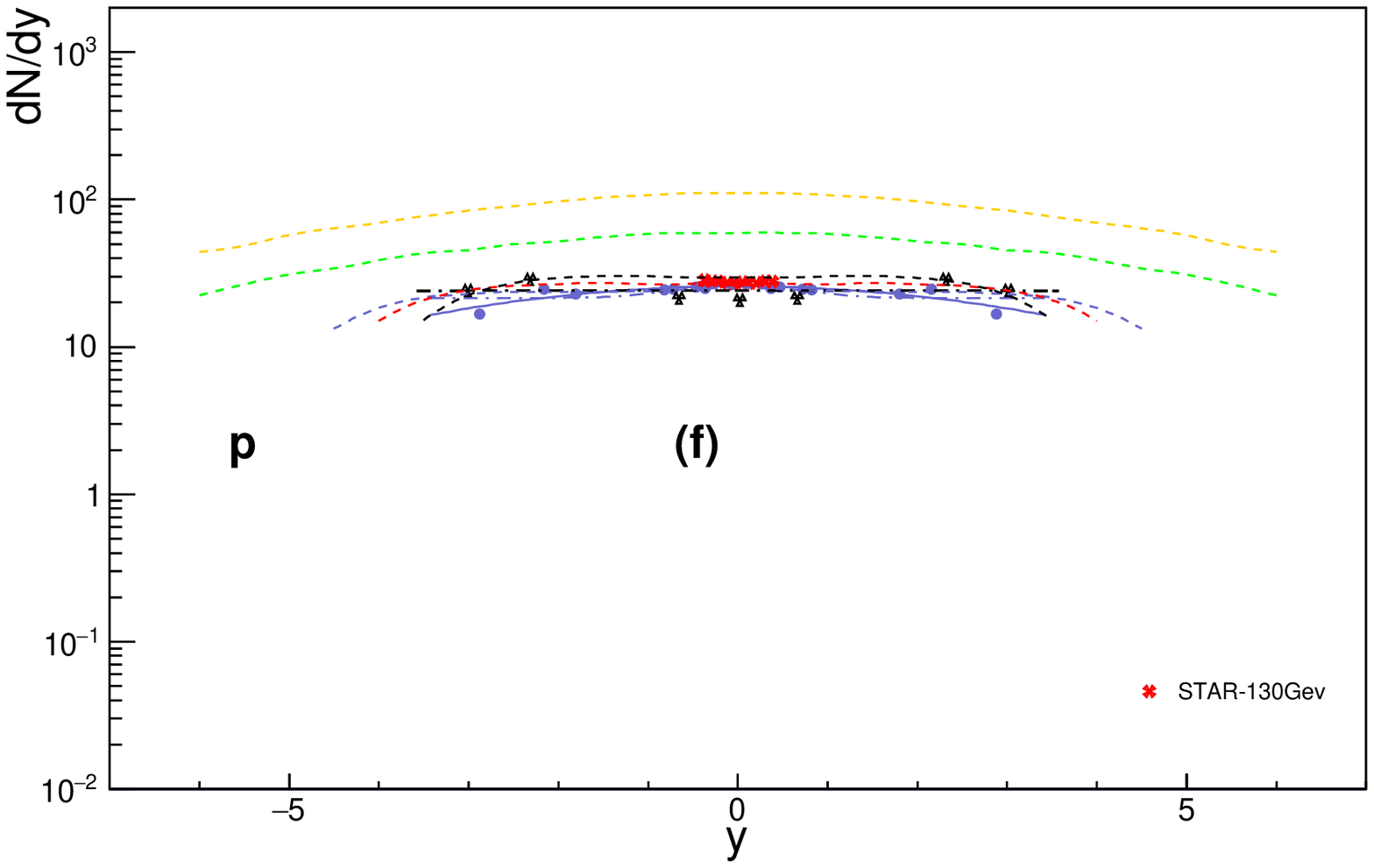} 
\includegraphics[width=7cm]{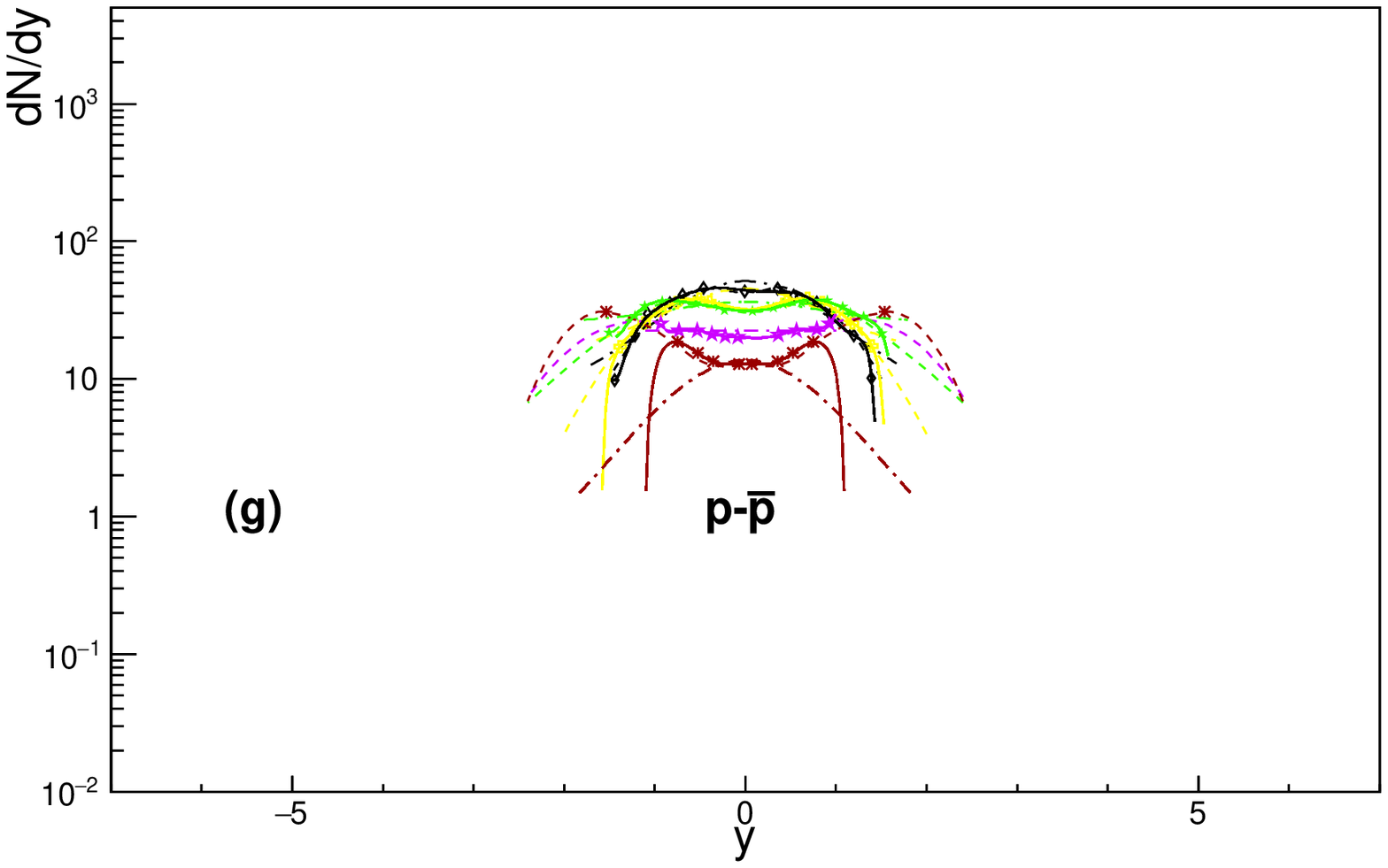} 
\caption{(Color online) In semi-log scale, the multiplicity per rapidity for $\pi^{-}$, $\pi^{+}$, $k^{-}$, $k^{+}$, $\bar{p}$, $p$, and $p-\bar{p}$ particle yields is depicted as a function of the rapidity. The experimental results \cite{Lee:2004bx,Afanasiev:2002mx,Bearden:2002ib,Bratkovskaya:2017gxq,Adams:2003ve,Zhou:2009zzh,Hebeler:2010xb,Bearden:2004yx,Seyboth:2005rn} (symbols) are compared to the Cosmic Ray Monte-carlo (CRMC EPOSlhc) event generator (dashed curves), section \ref{sec:crmc}, and also fitted to the hadron resonance gas (HRG) approach (dash-dotted curves), Eq. (\ref{eq(15)}) or Eq. (\ref{eq.(cccc)}) and to the Gaussian distribution function (solid curve).}
\label{fig:one}
\end{figure}

Figure \ref{fig:one} shows  the multiplicity per rapidity $dN/dy$ versus $y$ in a semi-log scale. The experimental results for $\pi^{-}$, $\pi^{+}$, $k^{-}$, $k^{+}$, $\bar{p}$, $p$, and $p-\bar{p}$ \cite{Lee:2004bx,Afanasiev:2002mx,Bearden:2002ib,Bratkovskaya:2017gxq,Adams:2003ve,Zhou:2009zzh,Hebeler:2010xb,Bearden:2004yx,Seyboth:2005rn} (symbols) are compared with the CRMC EPOSlhc event generator \cite{Werner:2005jf,Pierog:2009zt} (dashed curves). At the Large Hadron Collider (LHC) energies, $2760$ and $5500~$GeV, we introduce CRMC predictions. To the authors best knowledge, there is not such measurements to depict and compare with. The experimental results are fitted to he hadron resonance gas (HRG) approach (dash-dotted curves), Eq. (\ref{eq(15)}) or Eq. (\ref{eq.(cccc)}) and to the Gaussian distribution function (solid curve). For a better comparison, we keep the same $dN/dy$- and $y$-scales in all panels devoted to the different particle yields. 
	
There is a general observation that for all particles when the energy decreases, especially from top RHIC down to low SPS, i.e. from $200$ down to $6.3~$GeV, when disregarding our CRMC-predictions at LHC energies, the statistical fits seem becoming better and better. $\bar{p}$ and $p-\bar{p}$ are exceptions. Their fits become worse with decreasing the energy. Also, we generally observe that the HRG model can excellently describe the experimental/simulation results up to a relative narrow range around mid-rapidity. For a wider $y$-range, the ability of the HRG model to reproduce the results deduced from the experiments and the simulations becomes more and more worse. These are generic observations, from which final conclusions can be drawn. The goodness of the statistical fits shall be estimated, quantitatively. We find that the results from the CRMC EPOSlhc event generator match well with the experimental results. Accordingly, we present predictions, at LHC energies. Also, we observe that $dN/dy$ for the net-proton $p-\bar{p}$ seems to have two peaks. This might be understood due to the binomial assumed for this particle yield. For $\pi^{-}$, $\pi^{+}$, $k^{-}$, $k^{+}$, $\bar{p}$, and $p$, the fit parameters $V$ and $m_{\mathtt{T}}$, i.e. the volume of the fireball and the mass of the particle, respectively, are listed in Tab. \ref{tab1}. For net-proton $p-\bar{p}$, the corresponding fit parameters as deduced from Eq. (\ref{eq.(ddd)}) are given in Tab. \ref{tab2}.

\begin {table}[htb]
\begin{adjustbox}{width=\columnwidth}
\begin{tabular}{|c|c|c|c|c|c|c|c|}
\hline 
 particle & $\sqrt{{S}_{NN}} GeV$ & $a_{0}$ & $a_{1}$ & $a_{2}$ & $\chi^{2}$/dof & $V$ & $m_{T}$ \\ 
\hline 
  & 200  & $308.369\pm0.584$ & $-4.752\times10^{-3}\pm5.107\times10^{-3}$ & $2.341\pm 5.661\times10^{-3}$ &  0.888 & $2.536\times10^{5}\pm1.006\times10^{4}$  & $0.134\pm 0.067$  \\ 
\cline{2-8}
 & 62.4 &$226.487\pm4.722$  & $1.711\times10^{-3}\pm7.783\times10^{-2}$ & $1.778\pm 6.893\times10^{-2}$ & 120.8 & $2.938\times10^{5}\pm7.193\times10^{4}$ & $0.052\pm 0.026$  \\ 
\cline{2-8}
 &17.3  &$180.989\pm0.581$  &$-6.108\times10^{-6}\pm5.299\times10^{-3}$ &$1.436\pm 5.371\times10^{-3}$  & 0.568 & $3.188\times10^{5}\pm7.985\times10^{4}$ &  $0.035\pm 0.017$  \\ 
\cline{2-8}
 $\pi^{-}$&12.3  &$148.9\pm0.598$  & $-1.595\times10^{-3}\pm5.926\times10^{-3}$ & $1.279pm6.479\times10^{-3}$ & 0.518 &  $2.598\times10^{5}\pm5.951\times10^{4}$ & $0.041\pm0.018$   \\ 
\cline{2-8}
 &8.8  & $115.246\pm0.999$ & $4.526\times10^{-4}\pm1.136\times10^{-2}$ &$1.135\pm 1.152\times10^{-2}$  &1.342 & $1.937\times10^{5}\pm3.076\times10^{4}$  & $0.018\pm 0.009$ \\ 
\cline{2-8}
 & 7.8  &$104.838\pm1.835$  &  $8.039\times10^{-3}\pm2.104\times10^{-2}$& $1.045\pm 2.086\times10^{-2}$ & 4.144 & $3.195\times10^{5}\pm5.827\times10^{4}$  &  $0.013\pm   0.008$   \\ 
\cline{2-8}
 &6.3  & $88.074\pm0.272$  & $4.068\times10^{-3}\pm3.567\times10^{-3}$   &$1.004\pm 3.554\times10^{-3}$ &0.086 &  $3.492\times10^{5}\pm5.019\times10^{4}$  &  $0.009\pm 0.005$  \\ 
\hline
  & 200 &  $298.692\pm1.479$ &$-2.327\times10^{-4}\pm1.348\times10^{-2}$ &$2.265\pm 1.445\times10^{-2}$ & 17.22 & $2.691\times10^{5}\pm4.3869\times10^{4}$ & $0.120\pm 0.026$   \\ 
\cline{2-8} 
 $\pi^{+}$&62.4  &$-490.308\pm0.025$  &$-0.051\pm0.03$ & $1.826\pm2.588$  & 31.15  &  $3.024\times10^{5}\pm7.809\times10^{4}$ & $0.053\pm0.025$ \\ 
\cline{2-8}
 &17.3  &$159.21\pm1.932$ & $2.579\times10^{-3}\pm2.28\times10^{-2}$ & $1.548\pm2.499 \times10^{-2}$ &27.441  &   $2.098\times10^{5}\pm3.376\times10^{4}$  & $0.068\pm0.017$ \\ 
\hline 
  &200  &  $907.015\pm12.609$   & $3.379\times10^{-3}\pm0.033$ &$2.066\pm3.294\times 10^{-2}$ &34.44  & $1.477\times10^{5}\pm2.845\times10^{4}$ & $0.1210\pm0.0329$ \\ 
\cline{2-8} 
 &62.4  & $358.495\pm8.401$  &$2.049\times10^{-3}\pm4.630\times10^{-2}$ &$1.570\pm 3.731\times10^{-2}$  &12.304  &  $1.587\times10^{5}\pm2.315\times10^{4}$ &$0.026\pm 0.009$ \\ 
\cline{2-8} 
 &17.3  &$17.505\pm0.353$   & $1.311\times10^{-3}\pm1.95\times10^{-2}$ &$1.0673\pm 2.692\times10^{-2}$ & 0.472  & $2.922\times10^{5}\pm3.251\times10^{4}$  & $0.035\pm 0.009$   \\ 
\cline{2-8} 
 $\k^{-}$ & 12.3  &  $13.679\pm0.357$  &  $-0.119\pm1.867\times10^{-2}$& $0.833\pm 2.344\times10^{-2}$ &0.374  &$3.189\times10^{5}\pm4.360\times10^{4}$   & $0.009\pm 0.007$ \\ 
\cline{2-8}
 & 8.8  &$7.643\pm0.247$  & $-9.317\times10^{-2}\pm2.101\times10^{-2}$  & $0.769\pm 2.184\times10^{-2}$ &0.136  &  $2.656\times10^{5}\pm2.3145\times10^{4}$   &$0.002\pm 0.003$  \\ 
\cline{2-8}
 &7.8  & $6.499\pm0.130$  &$-8.229\times10^{-2}\pm1.530\times10^{-2}$ & $0.738\pm 1.586\times10^{-2}$  &0.057  &  $2.484\times10^{5}\pm1.480\times10^{4}$ &$0.003\pm 0.002$  \\ 
\cline{2-8}
 &6.3  & $5.961\pm0.185$  & $3.077\times10^{-3}\pm2.162\times10^{-2}$& $0.732\pm2.369 \times10^{-2}$&  0.097 &  $2.811\times10^{5}\pm1.169\times10^{4}$  & $0.003\pm0.001$ \\ 
\hline 
 &200  & $187.765\pm0.926$ &$-2.151\times10^{-4}\pm1.539\times10^{-2}$ &$2.383\pm 1.731\times10^{-2}$ & 6.996 & $1.183\times10^{5}\pm1.929\times10^{4}$  &$0.197\pm0.0414$   \\ 
\cline{2-8}
 &62.4  & $222.355\pm119.846$  &$-0.926\pm55.266$  & $0.807\pm51.701$ & 33.128 & $1.348\times10^{5}\pm3.187\times10^{4}$   & $0.061\pm0.026$ \\ 
\cline{2-8}
 &17.3  &  $31.241\pm0.547$ & $1.127\times10^{-2}\pm2.458\times10^{-2}$ & $1.217\pm 3.665\times10^{-2}$ &1.736  &$3.489\times10^{5}\pm5.374\times10^{4}$ & $0.083\pm0.021  $ \\ 
\cline{2-8}
$k^{+}$ &12.3  & $25.626\pm0.364$  &$-1.253\times10^{-3}\pm1.743\times10^{-2}$ & $1.093\pm2.373\times10^{-2}$  & 0.695 &  $4.268\times10^{5}\pm4.493\times10^{4}$   & $0.043\pm0.009$  \\ 
\cline{2-8} 
 &8.8  &$21.825\pm0.4389$  & $-2.923\times10^{-2}\pm2.497\times10^{-2}$  & $1.1218\pm 2.821\times10^{-2}$ &1.003  &  $4.907\times10^{5}\pm5.905\times10^{4}$   &  $0.029\pm0.007$   \\ 
\cline{2-8} 
 &7.8  & $21.775\pm0.261$ &$1.206\times10^{-3}\pm9.304\times10^{-3}$ &$0.8289\pm1.197 \times10^{-2}$  &0.231  &$8.088\times10^{5}\pm2.585\times10^{4}$ &  $0.009\pm0.001$ \\ 
\cline{2-8} 
 &6.3  &  $17.649\pm0.407$  & $4.083\times10^{-2}\pm1.434\times10^{-2}$  & $0.665\pm 1.475\times10^{-2}$ &0.528  &  $8.996\times10^{5}\pm3.806\times10^{4}$  & $0.006\pm 0.001$ \\ 
\hline 
$ \bar{p}$ &200  & $18.564\pm0.432$ & $-2.735\times10^{-3}\pm6.479\times10^{-2}$&$1.937\pm7.527\times10^{-2}$ & 1.131 &  $1.336\times10^{5}\pm3.851\times10^{4}$  &$0.158\pm0.063$  \\ 
\cline{2-8} 
 &62.4  & $11.551\pm0.059$    & $-1.265\times10^{-4}\pm1.416\times10^{-2}$ & $1.326\pm1.149\times10^{-2}$& 0.017 &  $1.901\times10^{5}\pm1.986\times10^{4}$ & $0.021\pm0.008 $    \\ 
\hline 
 &200  & $25.742\pm0.539$  &$-8.487\times10^{-4}\pm0.165$ &   $3.617\pm0.3166$ & 2.433  & $3.923\times10^{5}\pm7.8515\times10^{4}$  & $1.363\pm0.322$    \\ 
\cline{2-8} 
$p$ &130  & $400.26\pm0.285$  &$0.984\pm0.313$ &$0.372\pm0.03$ & 34.071 & $7.507\times10^{5}\pm3.456\times10^{4}$ & $9.11\pm4.149$    \\ 
\cline{2-8} 
 &62.4  &  $24.232\pm1.565$   & $-4.802\times10^{-2}\pm14.652$ &  $21.231\pm16.460 $ &16.75  & $4.001\times10^{5}\pm3.587\times10^{4}$   &$15.015\pm13.465$  \\ 
\hline  
\end{tabular}
\end{adjustbox}
\caption {The fit parameters obtained from the HRG approach for rapidity distributions for $\pi^{-}$, $\pi^{+}$, $k^{-}$, $k^{+}$, $\bar{p}$ and $p$, at various energies.}  
\label{tab1}
\end {table}

\begin {table}[htb]
\begin{adjustbox}{width=\columnwidth}
\begin{tabular}{|c|c|c|c|c|c|c|c|c|}
\hline 
 particle & $\sqrt{{S}_{NN}} GeV$ & $c_{0}$ & $c_{1}$ & $c_{2}$ & $c_{3}$& $\chi^{2}$ /dof & $V$ & $m_{T}$ \\ 
\hline 
 & 17.3  & $12.853\pm0.762$ & $17.501\pm0.0726$ &$-17.938 \pm0.064$ &$-140.123\pm 0.583$  &4.454 & $2.955\times10^{5}\pm6.042\times10^{4}$   &  $15.579\pm31.851$     \\ 
\cline{2-9} 
 &12.3 & $20.002\pm2.420$ &$-1.913\pm0.035$ & $2.623\pm0.053$  &$18.509\pm1.834 $ &6.260 & $3.928\times10^{5}\pm1.546\times10^{4}$   &  $14.673\pm57.729$   \\ 
\cline{2-9} 
 $p-\bar{p}$ &8.8  & $31.264\pm0.579$ &$-1.786\pm0.564$  & $19.819\pm1.876$  & $17.579 \pm1.757$ & 4.687 & $1.229\times10^{5}\pm7.0573\times10^{4}$   & $0.566\pm0.392 $   \\ 
\cline{2-9}
 &7.8  & $32.135\pm1.551$  &$-2.197\pm0.254$  &$37.869\pm2.675$ & $10.992\pm1.565$ &3.156 &  $9.241\times10^{5}\pm2.020\times10^{4}$   & $0.039\pm0.016$  \\ 
\cline{2-9}
 &6.3 &   $44.269\pm2.185$ & $-7.632\pm0.754$  & $6.814\pm0.476$   &$39.223\pm2.869$ &7.367   &  $1.874\times10^{5}\pm2.684\times10^{4}$    & $0.009\pm0.006$  \\ 
\hline  
\end{tabular} 
\end{adjustbox}
\caption {The same as in Tab. \ref{tab1} but for net-proton $p-\bar{p}$.}
\label{tab2}
\end {table}

\begin{figure}[!htb]
\includegraphics[width=7cm]{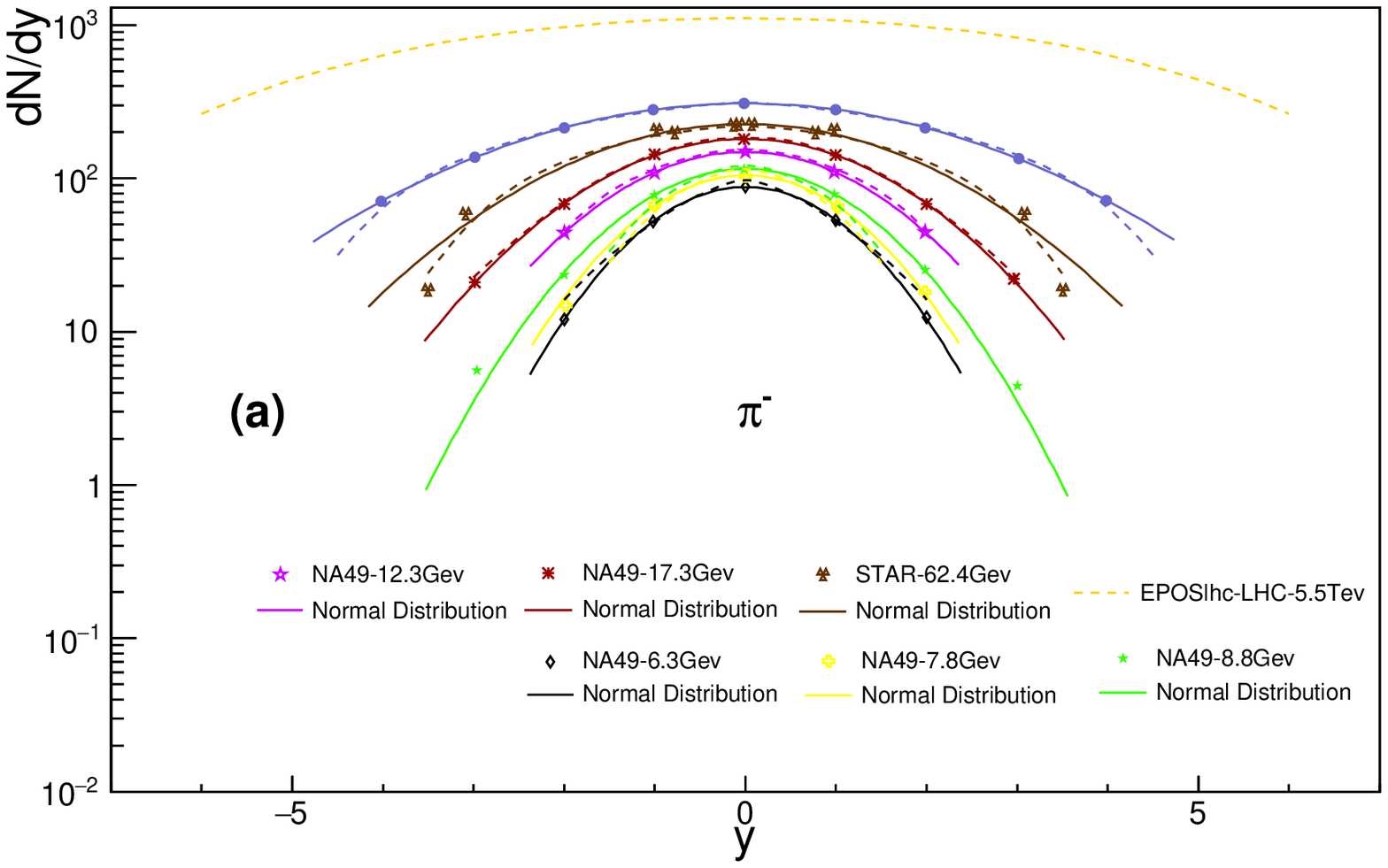}
\includegraphics[width=7cm]{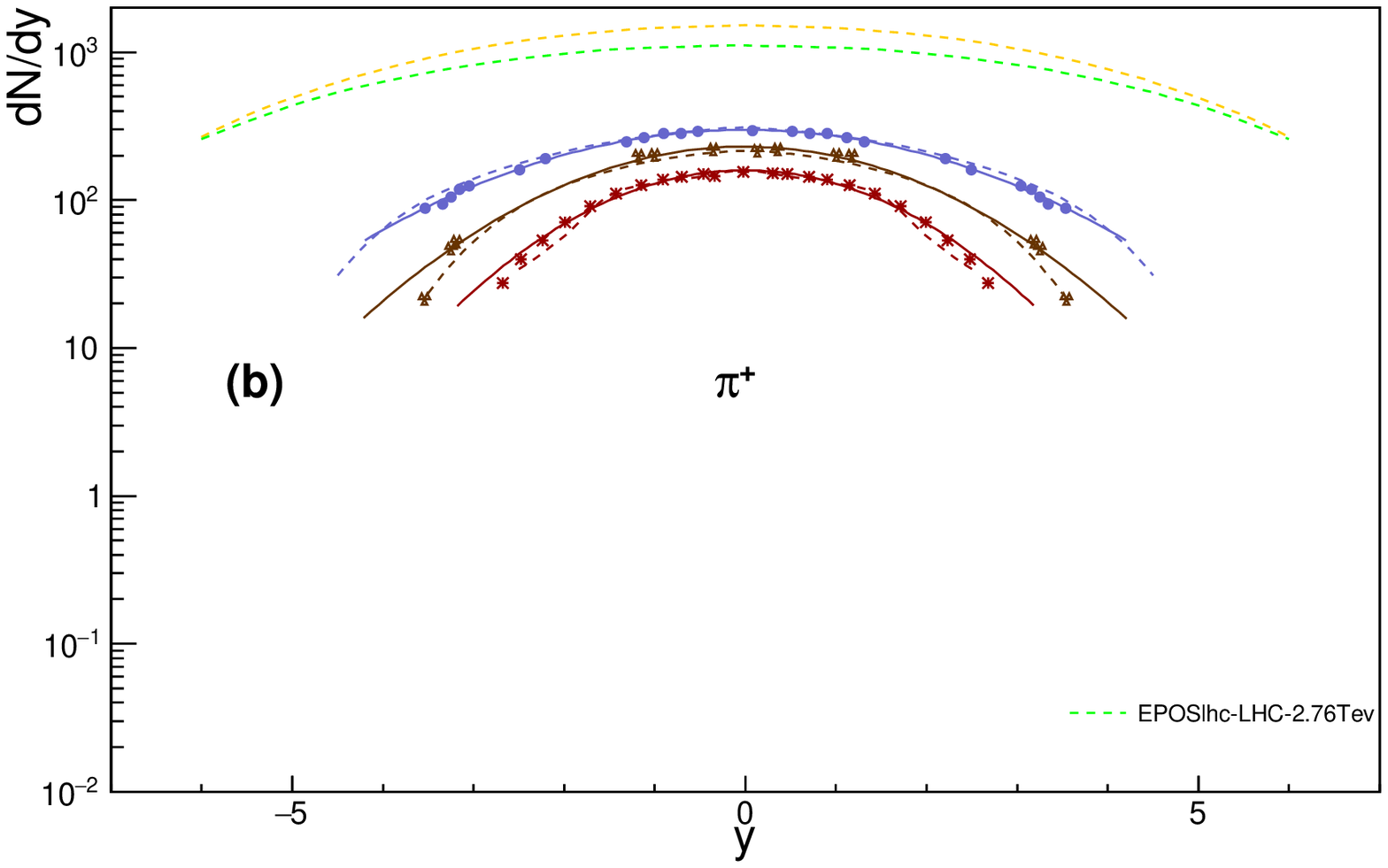} 
\includegraphics[width=7cm]{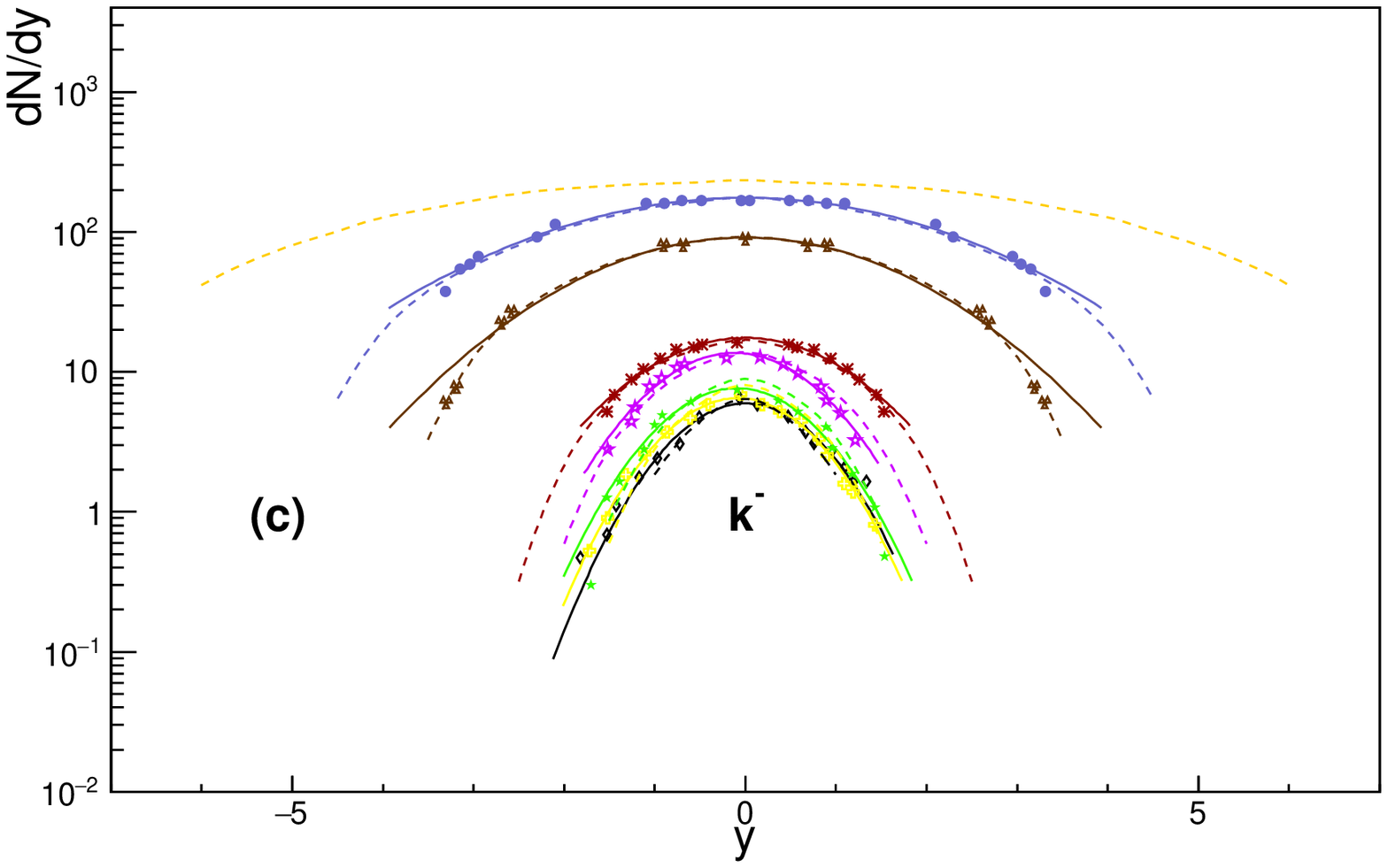} 
\includegraphics[width=7cm]{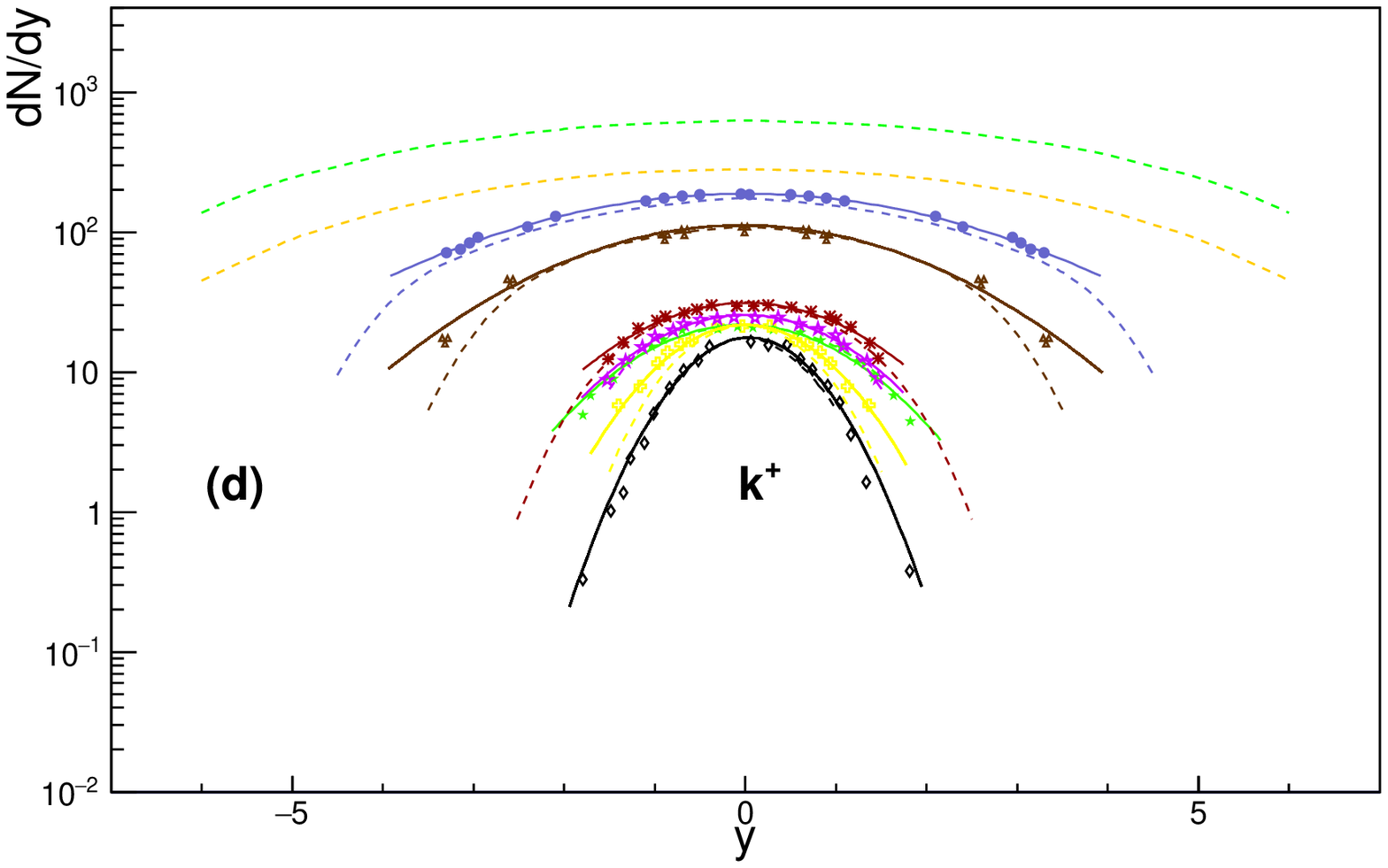}
\includegraphics[width=7cm]{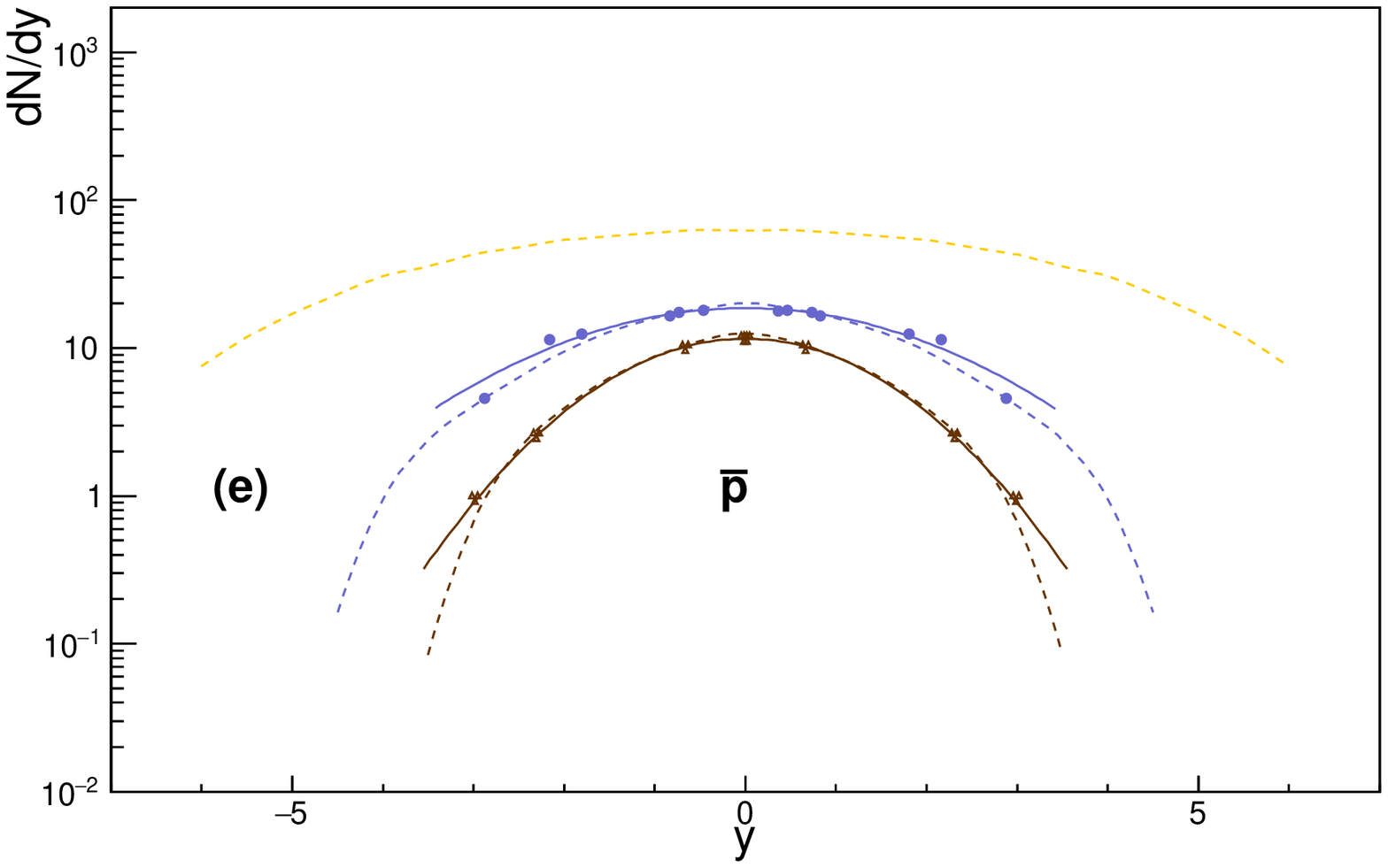} 
\includegraphics[width=7cm]{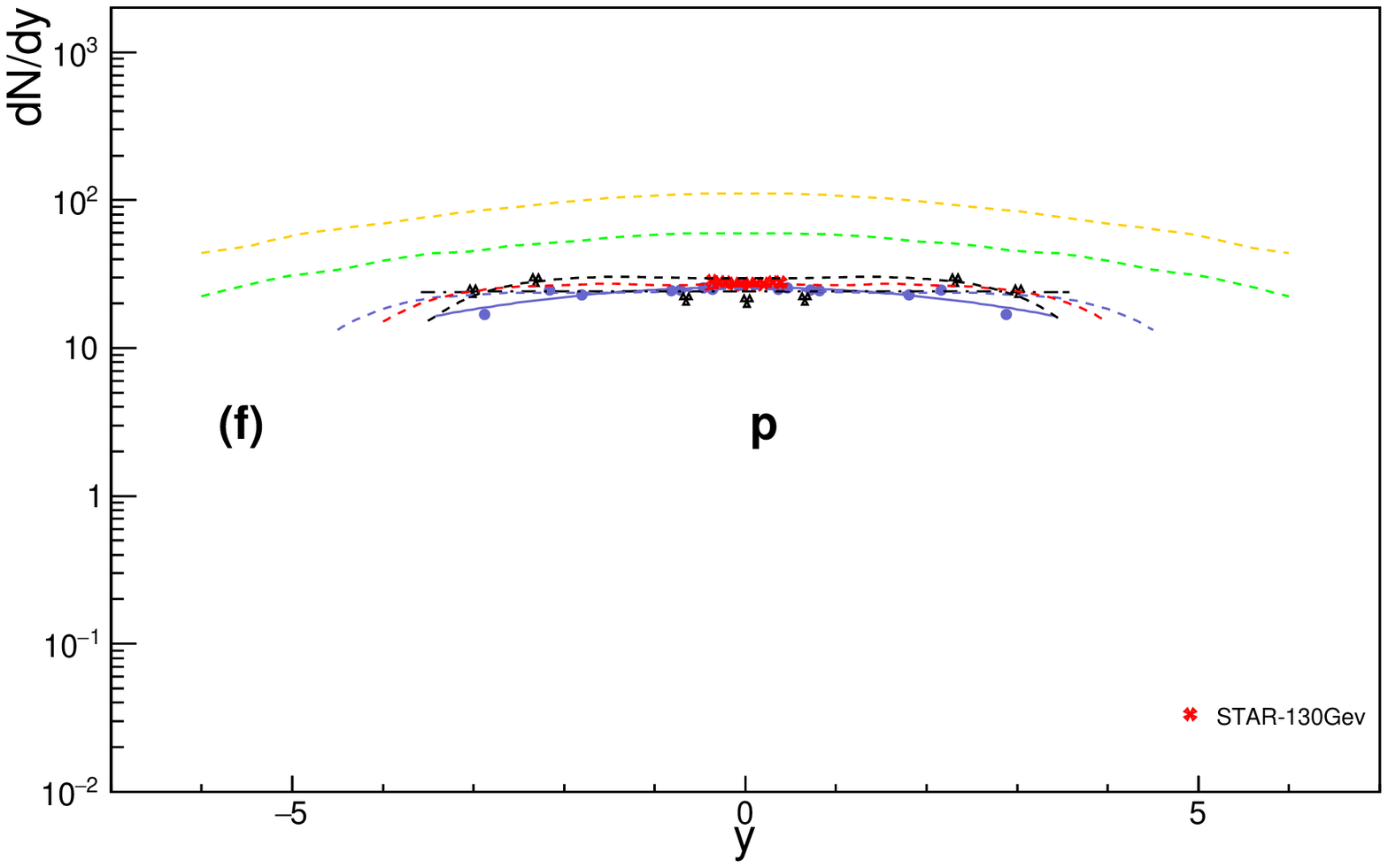} 
\includegraphics[width=7cm]{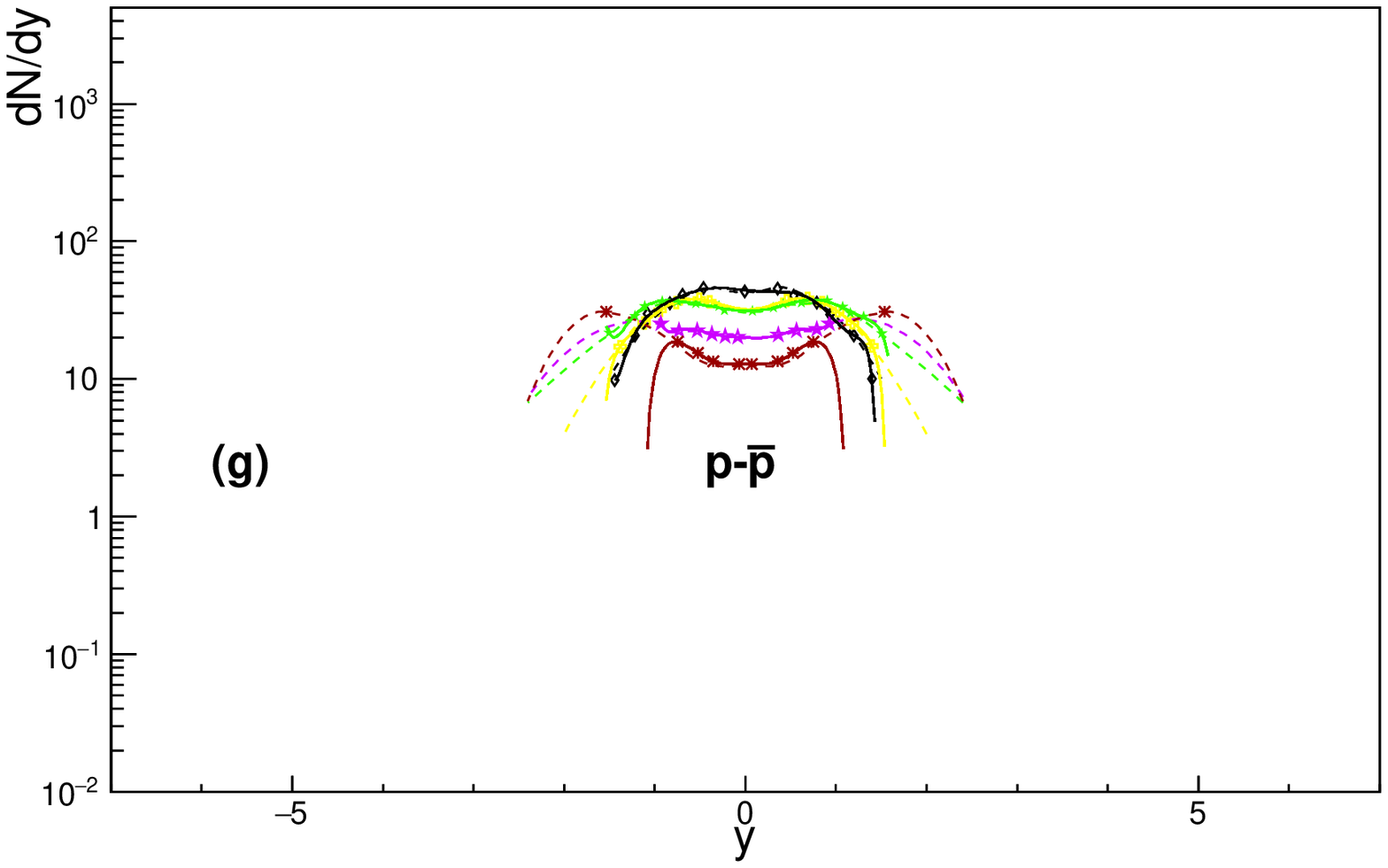} 
\caption{The same as in fig.(\ref{fig:one}) but here for Carruthers rapidity approach. }
\label{fig:three}
\end{figure}

Figure \ref{fig:three} shows $dN/dy$ versus $y$ from Caruthers rapidity approach, Eq. (\ref{eq(cc)}), and CRMC EPOSlhc event generator compared with the experimental data \cite{Lee:2004bx,Afanasiev:2002mx,Bearden:2002ib,Bratkovskaya:2017gxq,Adams:2003ve,Zhou:2009zzh,Hebeler:2010xb,Bearden:2004yx,Seyboth:2005rn}. We conclude that the results on $dN/dy$ excellently agree well with the CRMC EPOSlhc event generator.  The agreement with CRMC is also excellent. The goodness of corresponding fits is outlined in Tab. \ref{tab1}.

In light of this, we conclude that the correlations and fluctuations of the particle multiplicity as included in the Carruthers approach are essential for a better reproduction of the rapidity distributions of the various particles. The latter are likely isotropic and hence the overall results apparently match well with the Gaussian normal distribution. The corresponding parameters $\gamma_{\mathtt{2}}$, $\bar{y}^{n}$, and $\sigma_{\mathtt{m}}$ can be related to the resulting fit parameters, $\gamma_{\mathtt{2}}$= $a_{0}$, $\bar{y}^{n}$=$a_{1}$, and $\sigma_{\mathtt{m}}$=$a_{2}$, Eq. (\ref{eq.(llll)}).  

The multiplicity per rapidity of the well-identified particles $\pi^{-}$, $\pi^{+}$, $k^{-}$, $k^{+}$, $\bar{p}$, $p$, and $p-\bar{p}$ measured in various high-energy experiments, at energies ranging from $6.3$ to $5500~$GeV, are successfully compared to the Cosmic Ray Monte Carlo (CRMC) event-generator. The Carruthers and hadron resonance gas approaches are then fitted to both sets of results. We found that the Carruthers approach reproduces well the full range of multiplicity per rapidity for all produced particles, at the various energies, while the HRG approach fairly describes the results within a narrower rapidity-range.     

\section{Conclusions}
\label{conc}

We have calculated the multiplicity per rapidity $dN/dy$ for the well-identified hadrons $\pi^{-}$, $\pi^{+}$, $k^{-}$, $k^{+}$, $\bar{p}$, $p$, and $p-\bar{p}$ using two different approaches, namely HRG; a well-known framework based on thermal statistical assumptions and Carruthers approach based on correlations and fluctuations for hierarchy of the cumulant correlation functions and their linked-pair approximation, which in turn could be connected to negative binomial distributions and accordingly Gaussian- or exponential-like expressions for the rapidity distributions have been introduced.

The Carruthers and HRG approaches are then fitted to measurements at energies ranging from $6.3$ to $5500~$GeV and to corresponding simulations from the Cosmic Ray Monte Carlo (CRMC) event generator. The excellent agreement between the measurements and the simulations provides us with framework to compare between both approaches. We found that in the full range of rapidity, the multiplicity per rapidity is successfully reproduced in the Carruthers approach. The possible fluctuations and correlations as included in it seem to assure that the produced particles become isotropically distributed. On the other hand, the HRG approach restrictedly reproduces these anisotropic distributions. Accordingly, we conclude that the statistical assumptions alone - as included in the HRG approach - wouldn't be able to apply on a wide range of rapidity. Ingredients assuring fluctuations and correlations, such as flow and interactions, if integrated in the HRG approach would likely assure a better reproduction of the multiplicity per rapidity.

\section*{Acknowledgements}

The work of AT was supported by the ExtreMe Matter Institute (EMMI) at the GSI Helmholtz Centre for Heavy Ion Research.

%%%%%%%%%%%%%%%%%%%%%%%%%%%%%%%%%%%%%%
%%%   References
%%%%%%%%%%%%%%%%%%%%%%%%%%%%%%%%%%%%%%
\bibliographystyle{aip} 
\bibliography{CRMC_rapiditydensity5}

\end{document}